\documentclass[a4paper,12pt]{article}

\usepackage{latexstyle}
\usepackage{natbib}
\usepackage{todonotes}
\usepackage{hyperref}
\usepackage{subcaption}
\usepackage{lscape}
\usepackage{setspace}
\usepackage{caption}
\usepackage{xcolor}
\usepackage{float} 
\definecolor{salmon}{RGB}{246, 104, 95}
\definecolor{red4}{RGB}{118, 0, 2}
\definecolor{darkorange}{rgb}{0.8, 0.4, 0.0}
\definecolor{darkolivegreen}{rgb}{0.33, 0.42, 0.18}
\definecolor{darkolivegreen1}{rgb}{0.79, 1.0, 0.44}
\definecolor{forestgreen}{rgb}{0.13, 0.55, 0.13}
\usepackage{bigfoot}
\usepackage[font=small,labelfont=bf]{caption}

\title{Unveiling spatial patterns of population \\ in Italian municipalities
}

\author{Davide Fiaschi \thanks{University of Pisa,Via Ridolfi, 10, 56124 Pisa, Italy (e-mail: davide.fiaschi@unipi.it).}
 \and Angela Parenti\thanks{Angela Parenti (corresponding author), University of Pisa, Via Ridolfi, 10, 56124 Pisa, Italy (e-mail: angela.parenti@unipi.it).}
	\and Cristiano Ricci\thanks{Cristiano Ricci, University of Pisa, Via Ridolfi, 10, 56124 Pisa, Italy (e-mail: cristiano.ricci@ec.unipi.it).}}

\date{\today}

\begin{document}

\maketitle
\begin{abstract}
We study the evolution of population density across Italian municipalities on the based of their trajectories in the Moran space. We find evidence of spatial dynamical patterns of concentrated urban growth, urban sprawl, agglomeration, and depopulation. Over the long run, three distinct settlement systems emerge: urban, suburban, and rural. We discuss how estimating these demographic trends at the municipal level can help the design and validation of policies contrasting the socio-economic decline in specific Italian areas, as in the case of the Italian National Strategy for Inner Areas (Strategia Nazionale per le Aree Interne, SNAI).
\end{abstract}

\noindent \textbf{JEL Classification Numbers}: C14, O18, R12, R23

\noindent \textbf{Keywords}: random vector field, spatial dependence, spatial distribution dynamics, rural development, Moran space

\clearpage 
\tableofcontents

\newpage
\section{Introduction}

Nowadays, rural areas suffers by an increasing depopulation and degradation of their social texture \citep{copus2014territorial}. At the same time, urban areas encounter different challenges encompassing among others affordable housing, ageing population, inclusivity of migrants, social segregation, and climate change, generally related to increasing population density \citep{fioretti2020handbook}.
The essence of place-based policy lies not only in recognizing the heterogeneous challenges and opportunities of different areas but also in understanding the interplay of space and time that links each one. In this respect, the Sustainable Urban Development and the Rural Development strategies of the European Union have a spatial dimension that ranges from individual cities of various sizes, to different types of agglomerations encompassing multiple municipalities.
%Understanding the formation and evolution of these complex geographical areas requires overcoming the ``spatial mismatch'' between where people live and where job opportunities and services are located. 

In this paper, we provide a tool for estimating the joint dynamics of population density of municipalities and their Local Labour Areas (LLAs), where the use of the LLA allows us to take into account the spatial and economic interlinks that characterize the formation of urban/rural areas \citep{lamorgese2019stylized,manzoli2019house}. In particular, the evolution of the municipalities and their LLAs is estimated by a methodology inspired by the distribution dynamics proposed by \citet{quah1996twin, quah1997empirics}. The final goal is to forecast the population density at municipal level, which represents the basic information in the elaboration of several place-based policies.

From a methodological perspective, we advance the distribution dynamics by explicitly embedding the spatial dimension in the estimation of the dynamics, as in  \citet{fiaschi2014local, fiaschi2018spatial}. The technique is based on coupling each observation of the variable of interest with that of its spatial neighbours, i.e. considering it in the \textit{Moran space}, and then estimating the joint dynamics non-parametrically. The outcome of the estimation is an expected direction (represented by an arrow) at every point of the Moran space, that best approximates the trajectory followed by any unit around that point. Analogously to the classical estimation of distribution dynamics, this methodology allows the computation of the ergodic distribution, although keeping into account the spatial dependence. 
We depart from \cite{gerolimetto2022distribution} by directly including the value of the neighbour units in the dynamics instead of using an estimator that corrects for the potential bias derived from the omission of variables that are spatially correlated.

The estimate of the spatial dynamics of municipal population density from 1984 to 2019 in Italy highlights a strong heterogeneity across the different regions of the Moran space, with patterns of concentrated urban growth, urban sprawl, agglomeration, and depopulation. In particular, in the quadrant where both municipalities and their surrounding LLAs exhibit high population density (HH) (top-right in Moran space), we find a notable increase in the density of both municipalities and their LLAs, which suggests the presence of \textit{concentrated urban growth}. In this respect, it is not a surprise that the biggest 25 Italian municipalities belong to this region.
Conversely, in the quadrant characterized by high population density municipalities within low-density LLAs (HL) (bottom-right in Moran space), we only observe an increasing density in the LLAs. This points to the presence of \textit{urban sprawl}, with high-density municipalities spreading into less densely populated areas.
In contrast, in the quadrant where low population density municipalities are situated within high-density LLAs (LH) (top-left quadrant in Moran space), the primary significant pattern suggests an \textit{agglomeration process}. This means that there is a decreasing average density within the LLAs themselves, accompanied by an increase in the density of  some municipalities. In other words, populations tend to cluster in specific municipalities, while the surrounding municipalities experience a decrease in population density.
Finally, in the quadrant with both low-density municipalities and LLAs (LL) (bottom-left in Moran space), there is a tendency towards \textit{depopulation} with both the municipalities and their neighbours experiencing a decline in population. 

The estimate also reveals the emergence of a spatial structure with three types of \textit{human settlement systems}, i.e. the presence of three local spatial (long-run) equilibria (\textit{attractors}), in the Moran space: an \textit{urban attractor} characterized by high population density in both municipalities and their LLAs; a \textit{suburban attractor} with elevated density in LLAs compared to municipalities; and, a \textit{rural attractor} with low population density in both municipalities and their LLAs. 
The analysis of the basins of attraction suggests that approximately 65\% of all municipalities and 93\% of the total population should converge to the urban attractor, about 14\% of municipalities and 2\% of the population to the suburban attractor, and about 21\% of municipalities and 5\% of the total population to rural attractor.

Finally, our methodology is able to forecast the joint dynamics of each municipality and its surrounding LLAs, enhancing the precision of some targeted place-based policies. This is especially relevant to initiatives like the National Strategy for Inner Areas, as implemented in Italy, where one of the primary objectives is to contrast the current depopulation of some Italian regions \citep{barca2014}.

The paper is structured as follows. Section \ref{sec:time} explains the data sources for the Italian population density and provides a first description of its spatial dynamics; Section \ref{sec:timeSpace} describes our methodology and the results of its application to Italy; Section \ref{sec:innerAreas} discusses how our methodology can help the design of the National Strategy for Inner Areas, and finally Section \ref{sec:conclusion} concludes.

\section{The municipal population and local labour areas in Italy}\label{sec:time}

In this section we first discuss the data sources for the analysis of the dynamics of municipal population in Italy, and the partition of Italian municipalities in commuting zones, i.e. local labour areas. We then highlight the most important spatial patterns of Italian municipal population density in the period 19824-2019.

\subsection{Data sources and sample selection}\label{sec:data}

The analysis of spatial agglomeration is conventionally carried out at functional urban areas (FUAs) or local labour areas (LLAs), i.e. using population density and travel-to-work flows (commuting zone) as key information. Following the literature, we therefore use the population density at municipal and LLA levels. Data on municipal population and the definition of Italian LLAs come from ISTAT (Italian National Institute of Statistics).\footnote{Data and codes are available at the link: \url{https://github.com/PRINtimeSpaceEconAct/RVF_estimate}.}
Since 1861 the census of the Italian population is repeated every 10 years. From 1982 ISTAT also provides inter-census estimates of the population at yearly frequency. These estimates are obtained by updating the census population using the data on births, deaths, and internal and external migrations collected by the municipalities. Since October 2018 instead, ISTAT has been yearly conducting a sample survey by collecting the main characteristics of the Italian resident population at the municipal level to be integrated with administrative sources to get a permanent census.
Thus, considering both the \textit{10-year census data}, the inter-census estimates, and the \textit{permanent census}, the yearly data are available over the period 1982-2021. To avoid anomalies induced by the COVID pandemic, we decide to end the analysis in 2019.  Moreover, since we are interested in testing the time homogeneity of the estimated spatial dynamics, it is convenient to consider 1984 as the first year in order to subdivide the time span of observations into four sub-periods of the same length.
Finally, in the period 1984-2019 the definition of some of the municipalities changed several times, either because the municipal territory has modified, or because two or more municipalities merged together. To make it possible to compare municipalities over time we homogenized them to their definition in 2019.

Since 1981 ISTAT also classifies Italian municipalities into LLAs, areas where the bulk of the labour force lives and works (\citealp{franconil2018istat}). LLAs are usually considered a good unit of observation of spatial agglomeration since they contain both the place of residence and the place of work of almost all the residents. The initial partition of municipalities into LLAs has been updated every 10 years, up to 2001. In 2011 ISTAT radically changed the definition of Italian LLAs to be consistent with the Eurostat rules. For comparability, also the 2001 partition has been revised using the updated rule, while no update has been done for the partition of 1991 and 1981. 
Therefore, we only use the partition into LLAs of 2001 and 2011, updated with the latest Eurostat rules. Since the time span of our observation starts before the first updated partition in 2001, we adopt the partition of 2001 also for the years from 1984 up to 2001, and the partition of 2011 for the years from 2002 to 2019. We are aware that this could induce a bias in the representation of commuting patterns in years previous to the use of Eurostat rules, but only the partitions of 2001 and 2011 are fully comparable between themselves and based on the actual Eurostat rules. Finally, in Appendix \ref{app:changeLLA} we also control how much our estimates are affected by the change in the LLAs from 2001 and 2011.

\subsection{A first view on population density}\label{subsec:CS}

\begin{figure}[!htbp]
	\centering
	\begin{subfigure}[b]{0.47\textwidth}
		\centering
		\includegraphics[width=\textwidth]{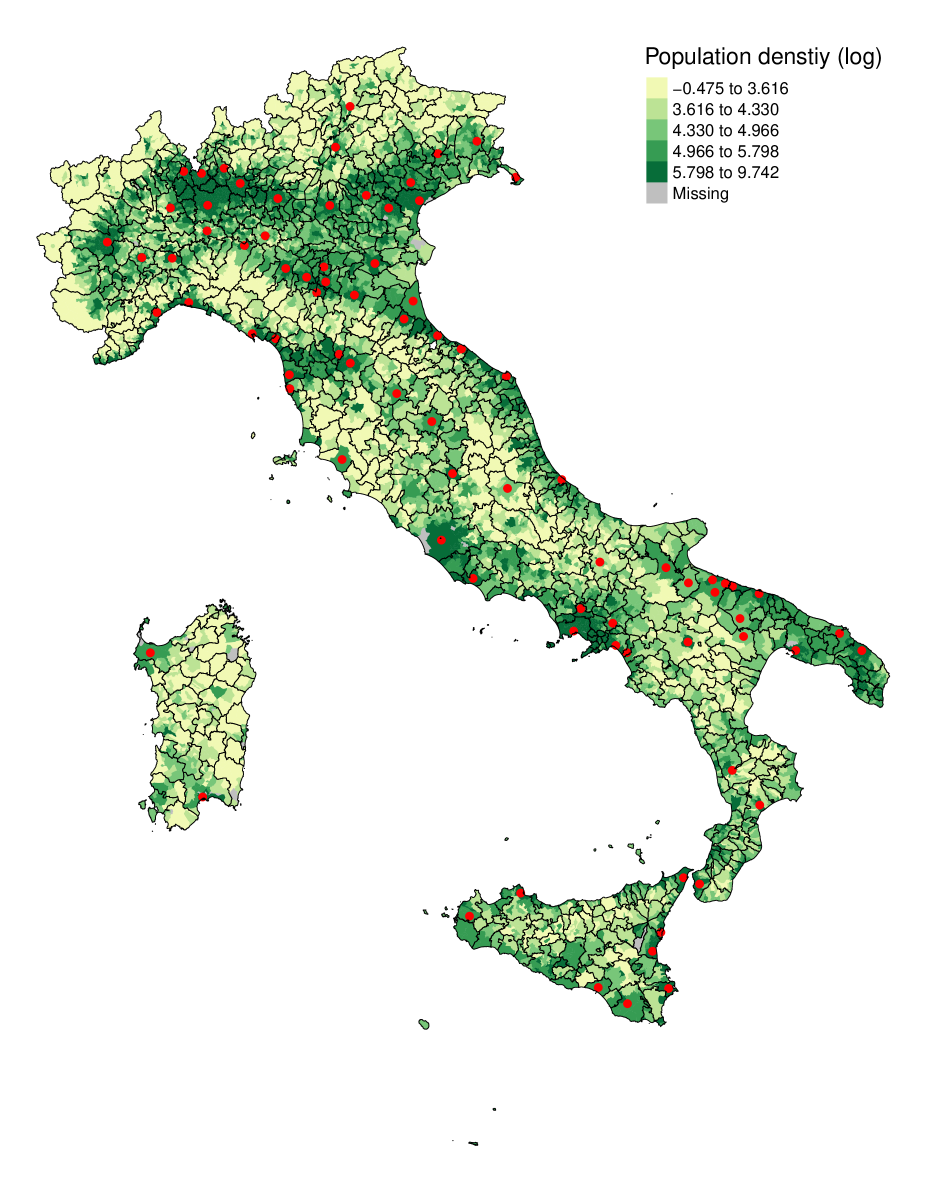}
		\caption{1984}
	\end{subfigure}
	\begin{subfigure}[b]{0.47\textwidth}
		\centering
		\includegraphics[width=\textwidth]{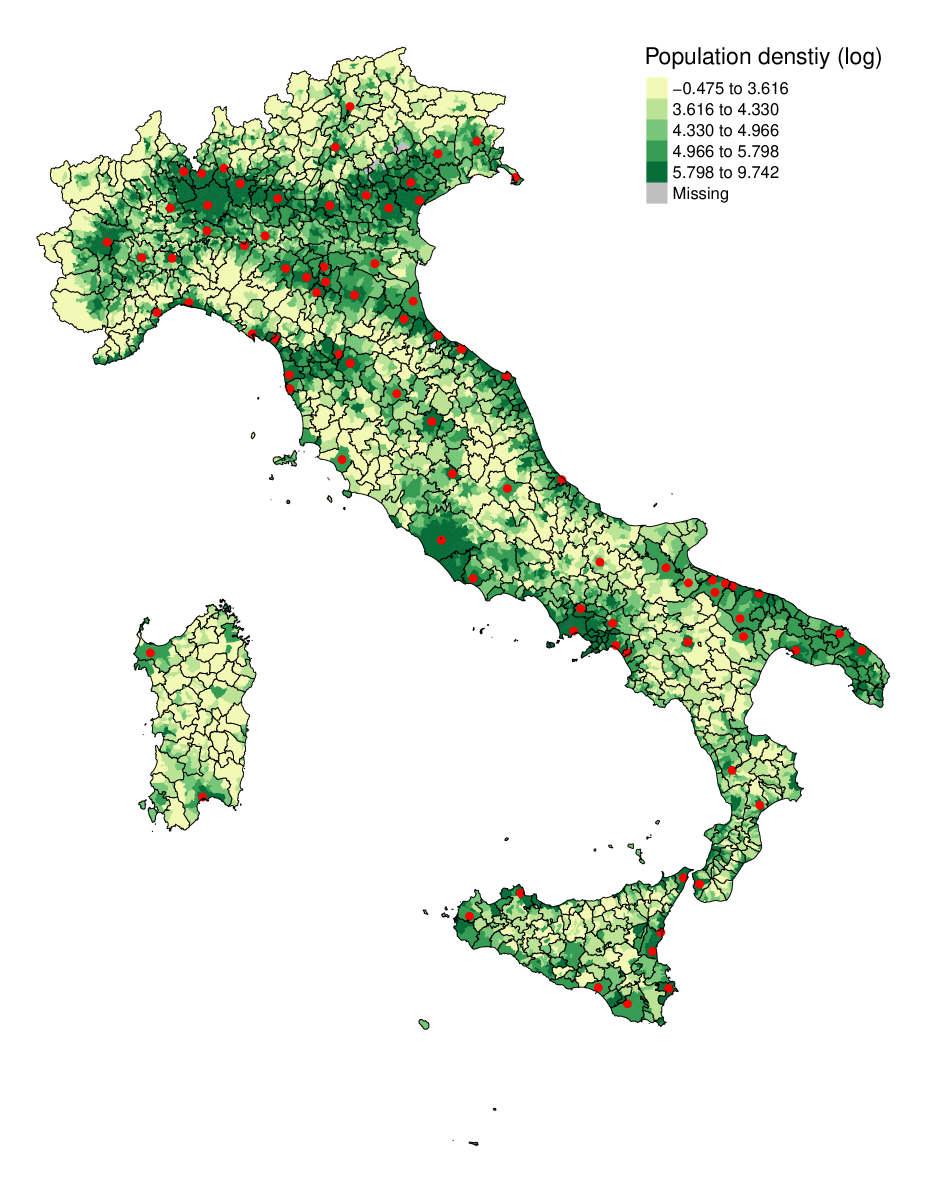}
		\caption{2019}
	\end{subfigure}
	\begin{subfigure}[b]{0.47\textwidth}
		\centering
		\includegraphics[width=\textwidth]{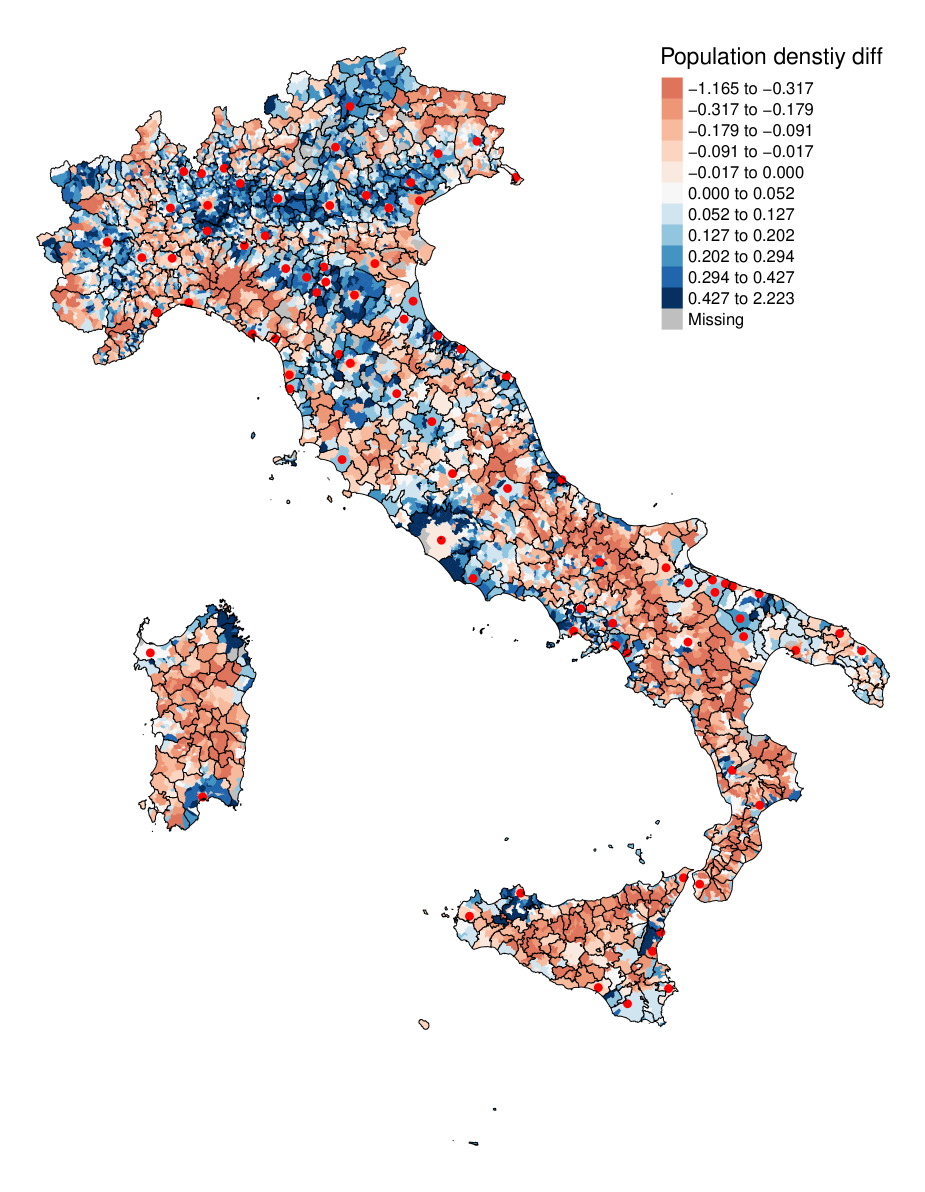}
		\caption{Difference between 1984 and 2019}
	\end{subfigure}
	\caption{The maps of population density of Italian municipalities in 1984 and 2019 (red points indicate the municipalities with more than 100k inhabitants) with LLAs boundaries. \\ \textit{Source: our elaborations on ISTAT data (Italian National Institute of Statistics).}}
	\label{fig:mapsMunicipalPopDensity}	
\end{figure}

Figure \ref{fig:mapsMunicipalPopDensity} reports the maps of the (log) population density of about 8000 Italian municipalities for the years 1984 and 2019, and its absolute variation over the period. A high level of spatial aggregation in population density can be noted, which seems to be not increasing over the period. The population density distribution appears strongly affected by geographical characteristics: the low density in the municipalities of the mountainous regions (along the Alps and Apennines) contrasts with the high density in the Po Valley and along the coasts. On the maps, red points indicate the municipalities with more than 100k inhabitants, among which we find the more restricted list of 14 Italian Metropolitan Cities as defined by Government (see Appendix \ref{tab:listMetroCities}). High-density municipalities seem to belong to more homogenous LLAs, and the same holds for low-density municipalities, while in the middle density municipalities the homogeneity is more faded. In the next Section \ref{sec:timeSpace}, Moran scatterplot will add support to this  evidence.
The time variations of population density confirm the presence of increasing spatial concentration, with a tendency for a raise in the population density of the already high-density municipalities and an opposite trend for the low-density ones. However, a more careful examination of the variations highlights that the general increase in the density of the main urban areas is associated with a decrease in the density of their surrounding municipalities, i.e., agglomeration. Nevertheless, there are some metropolitan cities (e.g., Turin, Milan, Rome, Naples, and Cagliari) where we can observe the opposite phenomenon, i.e., urban sprawl. Example of concentrated urban growth areas seems to be found along the borders of Po Valley, while areas of depopulation are mainly in correspondence of mountains.
\begin{figure}[!htbp]
	\begin{subfigure}[t]{0.32\textwidth}
		\includegraphics[width=\textwidth]{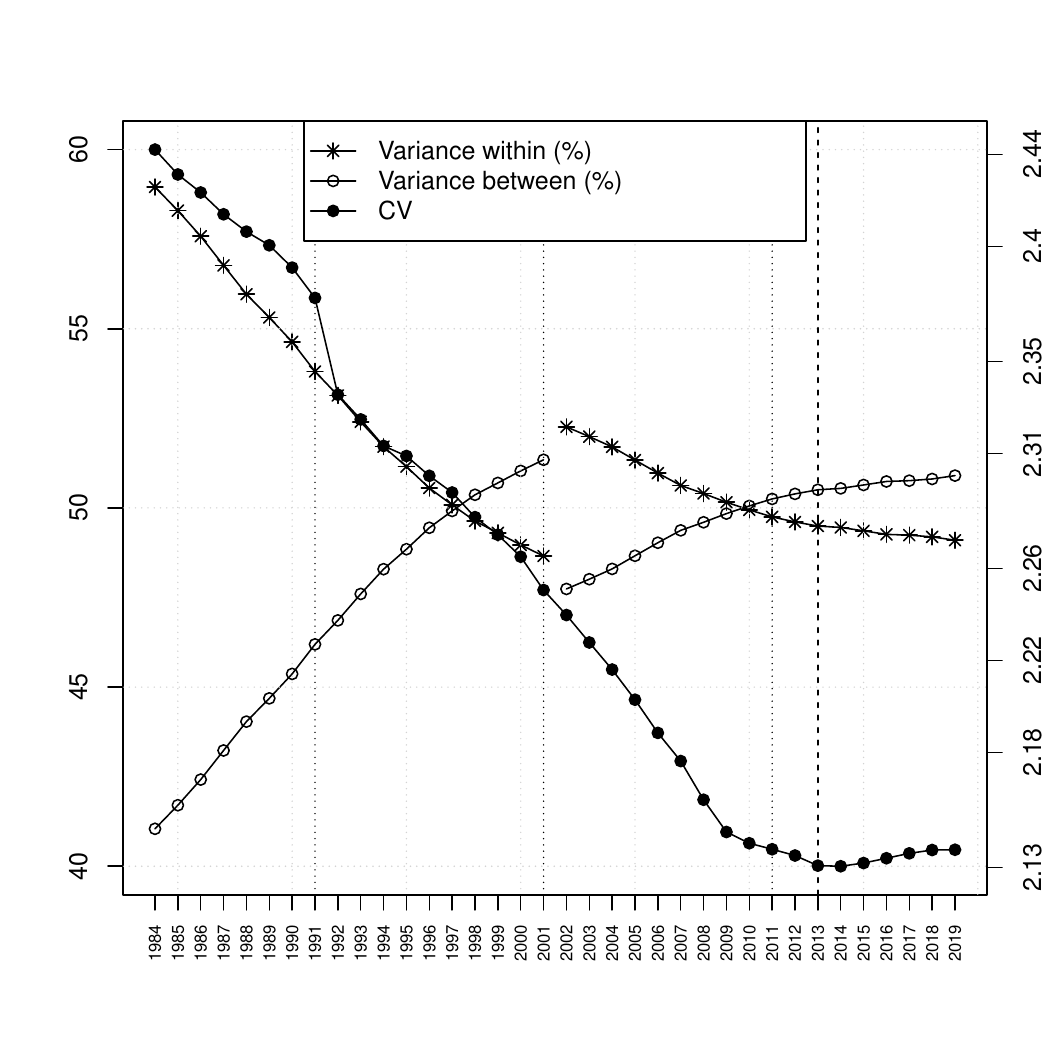}
		\caption{Coefficient of variation of municipal population density (scale on the right), variance between and within in percentage of total variance (scale on left).}
		\label{fig:varDecomposition_CV}
	\end{subfigure}
	\begin{subfigure}[t]{0.32\textwidth}
		\includegraphics[width=\textwidth]{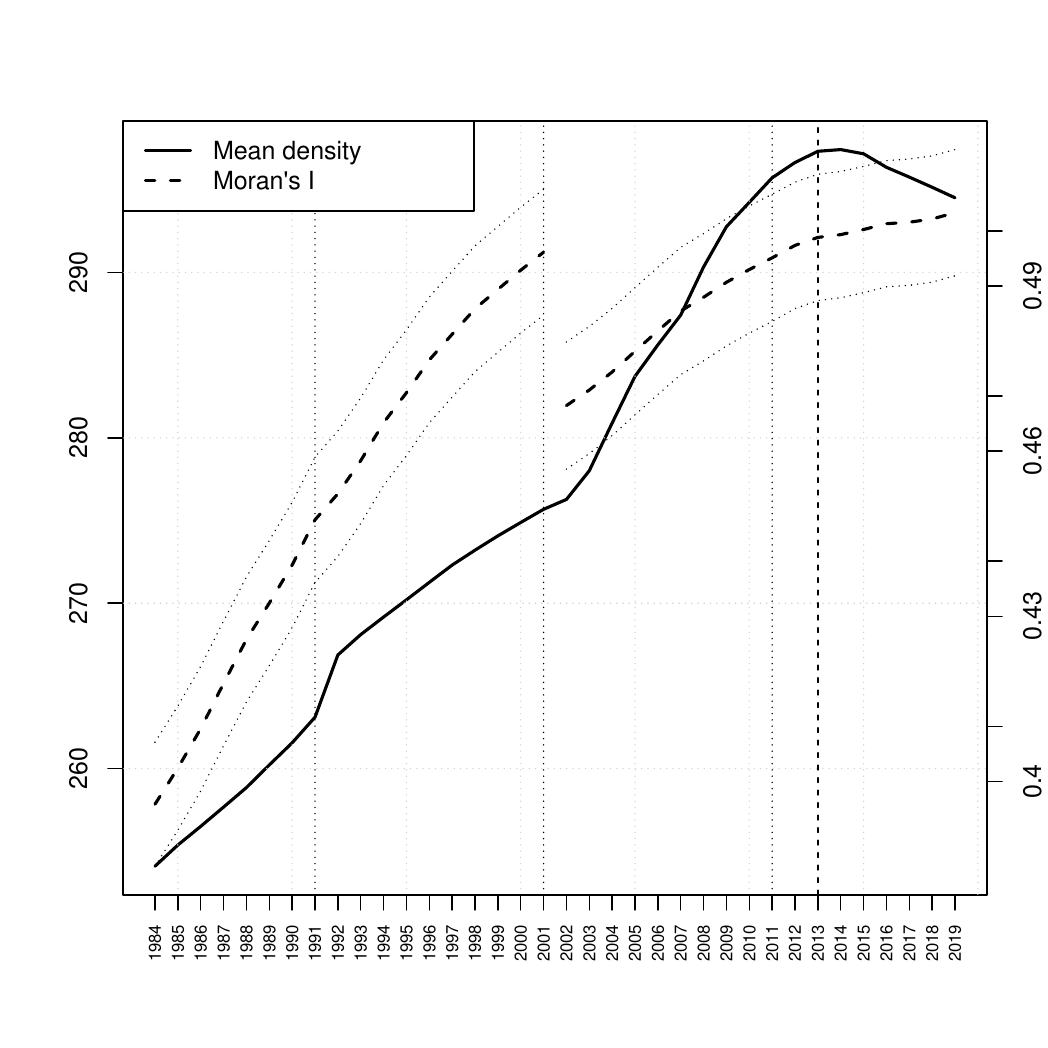}
		\caption{Mean density (scale on the left) and Moran's I index (scale on the right) of municipal population density, with its confidence bands at 95\%.}
		\label{fig:meanMoran}
	\end{subfigure}
	\begin{subfigure}[t]{0.32\textwidth}
		\includegraphics[width=\textwidth]{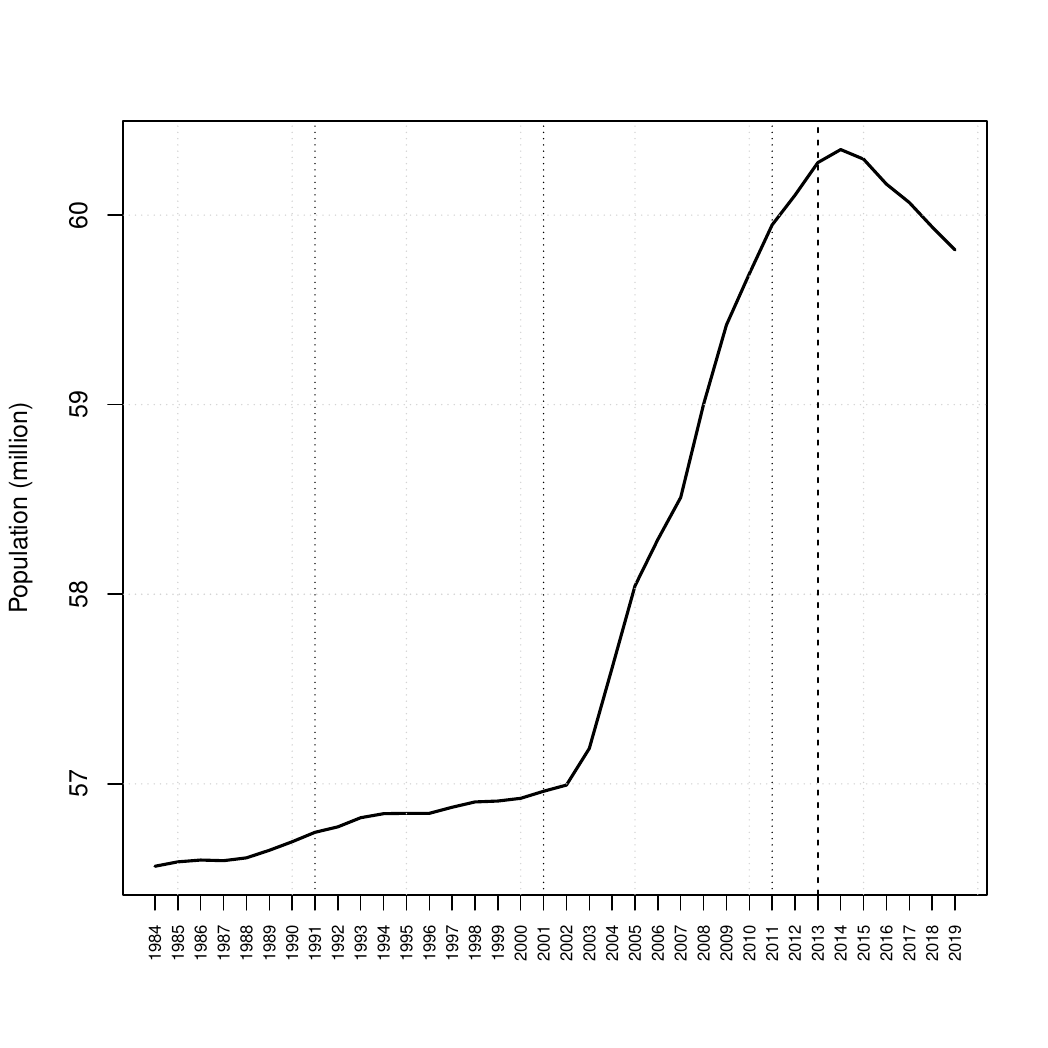}
		\caption{Italian population (millions).}
		\label{fig:totPop}
	\end{subfigure}
	\caption{Coefficient of variation (CV), variance between and within, mean density, Moran's I index and population of Italian municipalities in the period 1984-2019. Vertical dotted lines represent the years of Census. \\ \textit{Source: our elaborations on ISTAT data (Italian National Institute of Statistics).}}
\end{figure}
The dispersion among Italian municipal density is decreasing over time, as measured by the coefficient of variation reported in Figure \ref{fig:varDecomposition_CV}. However, evidence of agglomeration can also be grasped by Figures \ref{fig:varDecomposition_CV} and \ref{fig:meanMoran} which shows an increasing trend in the level of population density, at least until 2013, together with an increasing variance between LLAs and a raise in Moran's I index calculated on the base of the definition of LLAs, except for the downward jump in 2001 due to the change of LLAs definition. Such agglomeration is still more evident from the decreasing variance within LLAs.
As regards the mean density, its increasing tendency has reversed after 2013 for the decline in the total population (see Figure \ref{fig:totPop}); we also note a small upward jump between 1991 and 1992 likely induced by the inter-census estimates between the 1991 and 2001 censuses. The subsequent estimates do not seem to suffer from the same bias. No effect can be found in the time series of the total population, maybe due to the fact that the reconstruction of residents at the municipal level is more difficult.
The drop of the Moran's I in 2001 signals that the new definition of LLAs in 2001 tends to include high-density municipalities in LLAs with less populated municipalities. Typically, this can happen when the commuting zone of a municipality with high and increasing population, e.g. a metropolitan area, extends its borders to include some neighbouring municipalities, which before were member of less populated commuting zones.
The change in 2001, however, has not affected the increasing but concave shape of the curve of Moran's I index over time.

\section{The evolution of the population density\label{sec:timeSpace}}

We now estimate the spatial distribution dynamics of population density of Italian municipalities. In Section \ref{sec:MoranScatterPlot}, we first present the Moran scatterplot for 1984 and 2019, and then in Section \ref{sec:RVFestimation} the spatial distribution dynamics is estimated by a Random Vector Field (RVF) on the Moran space.

\subsection{The Moran scatterplot in 1984 and 2019 \label{sec:MoranScatterPlot}}
\begin{figure}[!htbp]
	\centering
	\begin{subfigure}[b]{0.47\textwidth}
		\centering
		\includegraphics[width=\textwidth]{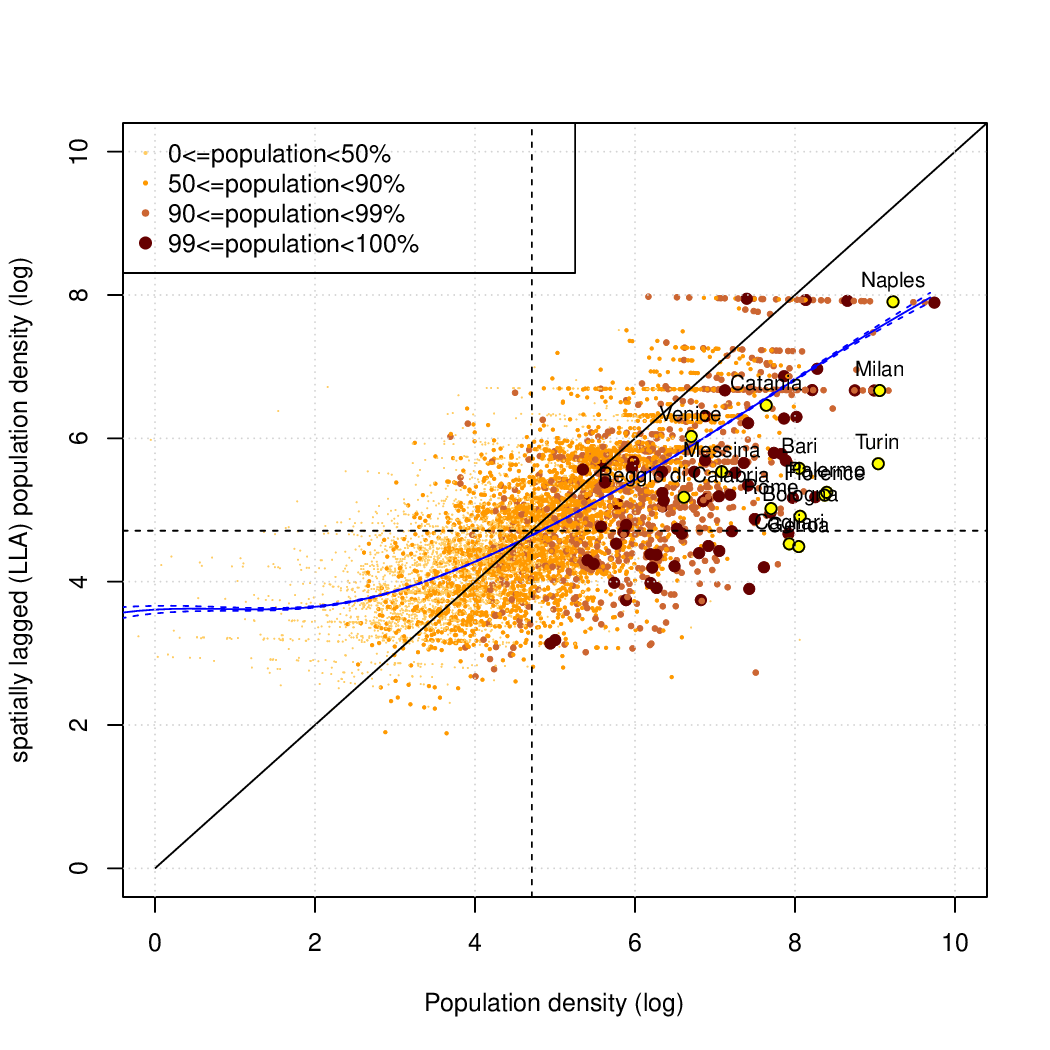}
		\caption{1984}
  \label{fig:moranScatterplot1984}
	\end{subfigure}
	\begin{subfigure}[b]{0.47\textwidth}
		\centering
		\includegraphics[width=\textwidth]{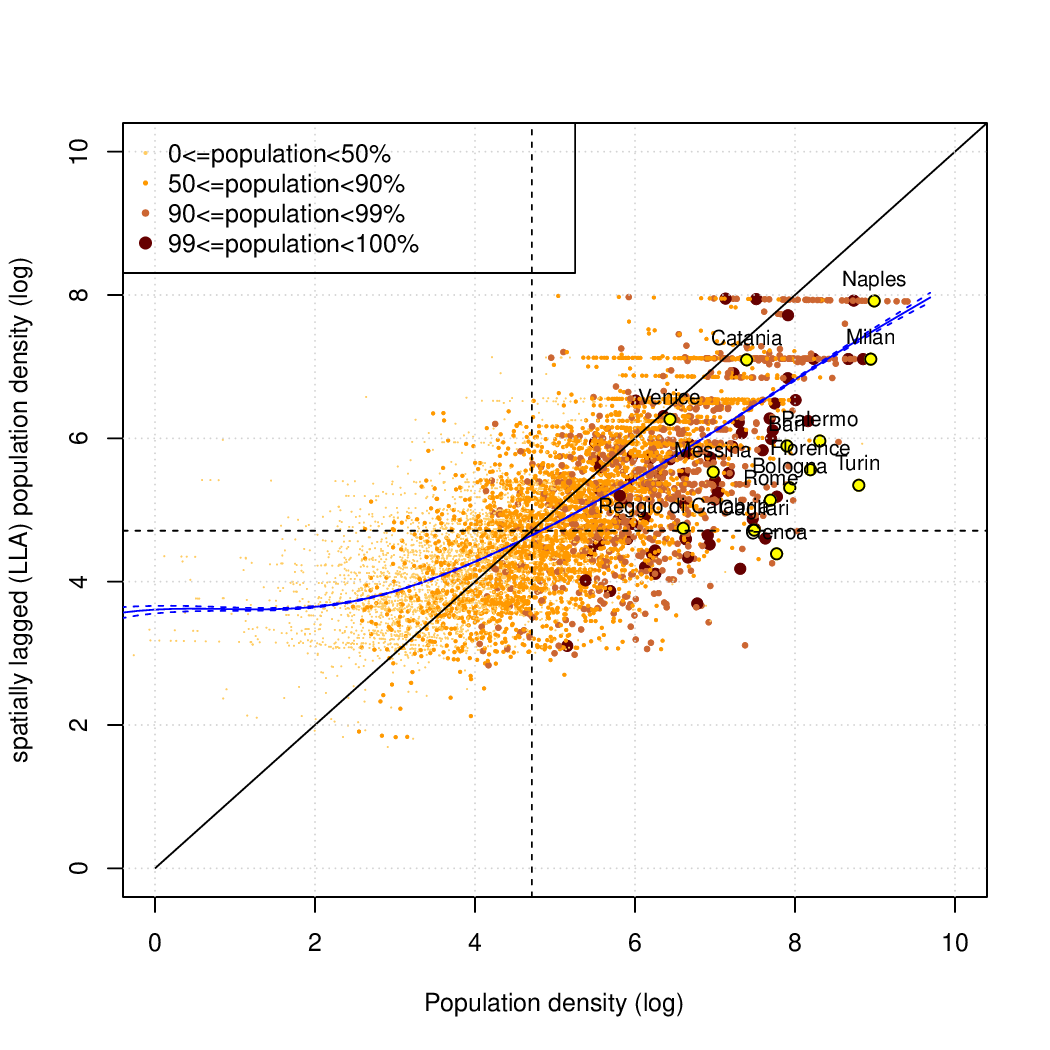}
		\caption{2019}
	\end{subfigure}
   \label{fig:moranScatterplot2019}
	%\caption*{\textit{Source: ISTAT (Italian National Institute of Statistics).}}	
	\caption{The Moran scatterplot of population density of Italian Municipalities in 1984 and 2019. The points represent the (log) population density of Italian municipality and the corresponding average value within its LLA. The size of the bullets is proportional to the municipality's total population (from bottom 50\%, represented by light yellow, to top 1\%, represented by brown points); the yellow points correspond to the 14 Italian metropolitan cities; the dashed lines are traced in correspondence of the average values for each axis. Blue lines is the estimated non-parametric Moran and its 95\% confidence bands.\\ \textit{Source: our elaborations on ISTAT data (Italian National Institute of Statistics).}}
	\label{fig:moranPopDensity}	
\end{figure}

Figure \ref{fig:moranPopDensity} shows the Moran scatterplot in 1984 and 2019. Each point represents the (log) population density of one municipality and its corresponding average density within its LLA. The size of the bullets is proportional to the municipality's total population. By yellow points we indicate the 14 Italian metropolitan cities, while the dashed lines are the average values for each of the two variables.  All the more populated municipalities (top 1\%, represented by brown points) show a density that is higher than the average. In contrast, the least populated municipalities (bottom 50\%, represented by clearer yellow) have a density lower than the average.\footnote{In most cases LLAs are made of municipalities that lie on the same horizontal line because the average population density of an LLA is roughly the same as the average of all the neighbours of each municipality that composes the LLA.} 

In the Moran space, we also report the non-parametric estimation of Moran's I with its 95\% confidence bands (blue solid and dashed lines, respectively). More precisely, the local Moran's I index corresponds to the local slope of the estimated curve reported in the Figure \ref{fig:moranPopDensity}.\footnote{The non-parametric Moran is estimated using the \verb|sm| R library.}
In 1984 we observe a significant positive spatial dependence for the whole range of population density, except for low-density municipalities, where such dependence is almost zero. Moreover, on the HH quadrant, the non-parametric Moran is below the 45-degree line while on the LL is above, i.e. a municipality with a density higher than the average is surrounded by municipalities with high density but lower than its own; on the other hand, in the LL quadrant a municipality with a density lower than the average is surrounded by municipalities with a density lower than the average but higher than its own.

As expected, almost all the metropolitan cities are in the HH quadrant, below the bisector and the non-parametric Moran. Exceptions are Venice, Catania, and Naples which are much nearer to the bisector, meaning that they are closer in density to their neighbours. This may be explained by the geographical characteristics of their territory, which prevent their vertical extension (Venice is on a lagoon, and Catania and Naples are close to Etna and Vesuvius volcanoes, respectively).
The Moran scatterplot in 2019 looks very similar to the one of 1984, but Naples, Catania and Venice are closer to the bisector; the estimated Moran is higher in the HH quadrant; and finally, Genoa and Turin see to decrease the average density of their LLAs for the change in their LLA definition (see Appendix \ref{app:changeLLA} for more details).

\subsection{The estimation of spatial dynamics\label{sec:RVFestimation}}

The dynamics of the municipal population density can be estimated using the \textit{stochastic kernel} proposed by \cite{quah1996twin, quah1997empirics} that allows tracking intra-distribution dynamics of units as a generalization of Markov transition matrices to continuous state space. However, when analysing the spatial evolution of population density, it becomes essential to account for spatial dependence directly. This presents a challenge known as the ``curse of dimensionality'', as highlighted by \cite{silverman1986density}.\footnote{One attempt to avoid this issue is the conditioned stochastic kernel of \cite{quah1997empirics} where the variable of interest is taken in relative terms with respect to the neighbour level. Although this allows to take into account the spatial component, it is not able to trace the joint increase/decrease of both the municipal density and that of its neighbours when the relative level remains unchanged. Another approach is the one of \cite{gerolimetto2022distribution}, where the authors consider the spatial dependence arising from omitted variables that are spatially correlated. Although this procedure takes space into account in the estimation of the stochastic kernel, thus avoiding the bias from omitted autocorrelation, it is not able to track the dynamics of neighbours.}
Here, we use an alternative methodology based on the estimate of a RVF on a Moran space, which estimates the joint dynamics of the population density of the municipality and of its LLA as a \textit{vector field}.

\subsubsection{The estimation of the spatial dynamics by a RVF on a Moran space}
Let $\mathbf{y}^{t_0}=(y_j^{t_0})$ and $\mathbf{y}^{t_0+\tau}=(y_j^{t_0+\tau})$, for $j=1,\dots,N$, be the observed level of (log) population density for municipality $j$ at year $t_0$ and $t_0+\tau$, respectively, and $N$ the number of observations in the sample. Moreover, let $\mathbf{W}$ the $N\times N$ the row-standardised spatial weight matrix defining for each municipality its membership to an LLA. Define $\mathbf{z}_j^{t_0}:=\left(y_j^{t_0}, (\mathbf{Wy}^{t_0})_j\right)$ and $\mathbf{z}_j^{t_0+\tau}:=\left(y_j^{t_0+\tau}, (\mathbf{Wy}^{t_0+\tau})_j\right)$ the municipal observations in Moran space in year $t_0$ and $t_0+\tau$, respectively. Therefore, we can define the observed movement as $\boldsymbol{\delta}_j^\tau:=\mathbf{z}_j^{t_0+\tau}-\mathbf{z}_j^{t_0}$. 
For each point $\mathbf{z}$ in the Moran space define the expected movement as the weighted average of all the observed ones around the point, as:
\begin{equation}
	\widehat{\mathbf{RVF}}_\tau(\mathbf{z}):=\sum_{j=1}^N \omega_j^{t_0}(\mathbf{z})\boldsymbol{\delta}_j^\tau
    \label{eq:RVF}
\end{equation}
where:
\begin{equation*}
    \omega_j^{t_0}(\mathbf{z}):= \dfrac{\dfrac{1}{Nh^2}K\left(\dfrac{\mathbf{z}-\mathbf{z}_j^{t_0}}{h}\right)}{\hat{f}_{Y^{t_0},WY^{t_0}}(\mathbf{z})},
\end{equation*}
\begin{equation*}
   \hat{f}_{Y^{t_0},WY^{t_0}}(\mathbf{z}):=\sum_{i=1}^N \dfrac{1}{Nh^2}K\left(\dfrac{\mathbf{z}-\mathbf{z}_i^{t_0}}{h}\right),
\end{equation*}
where $K(\cdot)$ is a \textit{kernel function} in $\RR^2$ and $h$ is a smoothing parameter (\textit{bandwidth}) \citep{silverman1986density}. 
Heuristically, Eq. \eqref{eq:RVF} represents the expected movement in point $\mathbf{z}$ obtained by summing up all the observed movements $\boldsymbol{\delta}_j^\tau$ from $t_0$ to $t_0+\tau$, with weights $\omega_j^{t_0}(\mathbf{z})$ depending on  the distance of each observed movements from $\mathbf{z}$ at time $t_0$. The latter is equal to the relative contribution of observation $j$ to the joint density estimation of observations $\mathbf{z}_i^{t_0}$ calculated at the evaluation point $\mathbf{z}$.
In the estimate we consider a more sophisticated version of Eq. \eqref{eq:RVF} which takes into account the covariance between the two components of $\mathbf{z}_j^{t_0}$ and a bandwidth which varies with respect to the local density of observations \citep{silverman1986density,fiaschi2018spatial}.\footnote{In particular, our estimation has been carried out with an \textit{Epanechnikov kernel} for $K\left(\cdot\right)$, $h=0.21$ and $\alpha=0.0067$ (the parameter adapting bandwidth to local density of observation), see Appendix \ref{app:RVF_flavour}.} We refer to Appendix \ref{app:RVF_flavour} for more technical details.
The main advantage of using RVF is that it attenuates the curse of dimensionality since it is only based on two-dimensional density estimation instead of four-dimensional one as in the case of stochastic kernel.\footnote{The estimation of the stochastic kernel keeping spatial dependence explicitly would require to estimate a joint non-parametric density of $\left(\mathbf{y}^{t_0}, \mathbf{Wy}^{t_0}, \mathbf{y}^{t_0+\tau}, \mathbf{Wy}^{t_0+\tau}\right)$.} In particular, the number of dimensions is reduced from four to two by considering the observed movements conditioned to the initial observation. The latter condense all the information on the dynamics. The estimated RVF can be also easily visualized by drawing an arrow for each evaluation point in the Moran space, as shown in Figure \ref{fig:RVF_84_19}.

\subsubsection{The estimate of RVF for Italian municipalities}

\begin{figure}[!htbp]
	\centering
	\begin{subfigure}[b]{1\textwidth}
		\centering
	\includegraphics[width=\textwidth]{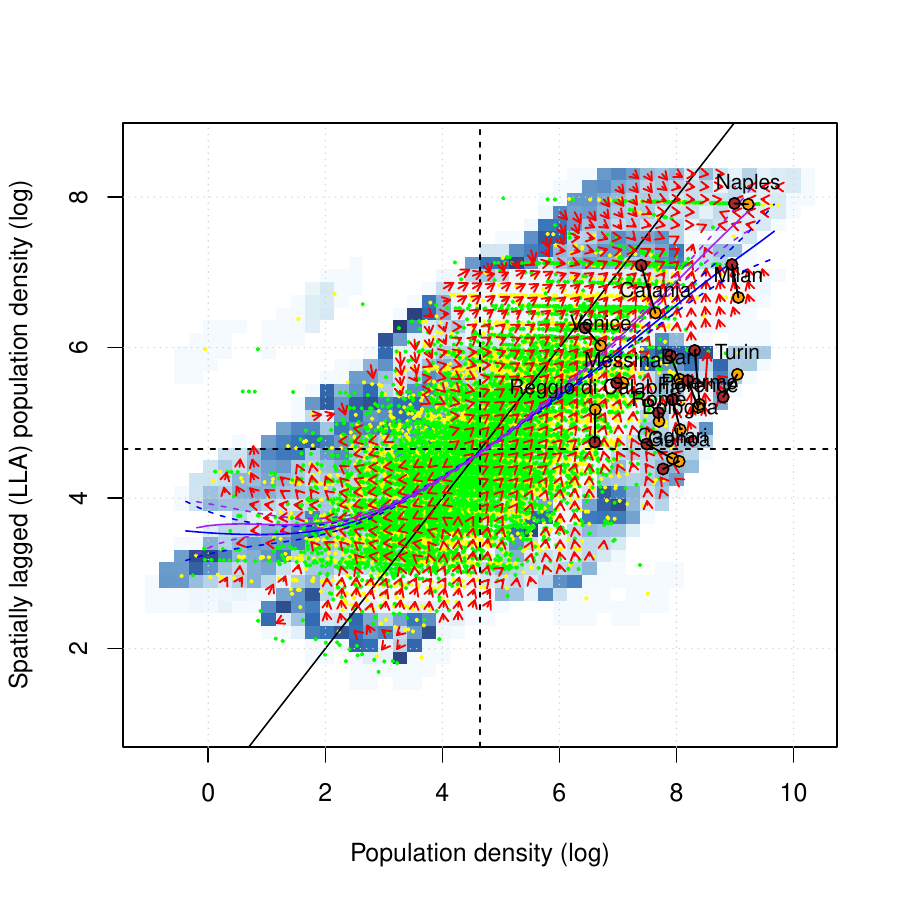}
	\end{subfigure}
		\caption{The estimate of RVF of Eq. \eqref{eq:RVF} in the Moran space for the period 1984-2019. Red arrows are the directions statistically significant at 5\% level (the length of each arrow has been divided by a factor 5 for graphical purposes).
		Estimated variance of the arrows' direction is reported in blue (the darker the colour the higher the variance). Observations in 1984 and 2019 are coloured in yellow and green, while metropolitan cities in orange and brown, respectively. Blue and purple lines are the estimated non-parametric Moran and their 95\% confidence bands in 1984 and 2019.
		\\ \textit{Source: our elaborations on ISTAT data (Italian National Institute of Statistics).}}
  \label{fig:RVF_84_19}
\end{figure}

In Figure \ref{fig:RVF_84_19} for each evaluation point we report in blue the variance of the arrows' direction (the darker the colour the higher the variance), and only the arrows that are significant at 5\% significance level.\footnote{The standard errors of the estimated arrows are derived from non-parametric block bootstrap (\citealp{efron1986bootstrap}).} Observations in 1984 and 2019 are coloured in yellow and green, while metropolitan cities in orange and brown, respectively. We also draw the estimated non-parametric Moran and its 95\% confidence bands in both years (blue and purple lines for 1984 and 2019, respectively). 

In Figure \ref{fig:RVF_84_19} most of the arrows in the HH quadrant point to upper right, i.e. towards an increase both in the level of municipal density and its LLA, which represents a pattern of concentrate urban growth. The exceptions are the municipalities with the highest density, as Naples, Milan and Turin, with a direction toward upper left, which suggests a pattern of urban sprawl. As result, the non-parametric Moran in quadrant HH displays a slight rotation toward the bisector. It is also possible to notice the presence of an attractor around the location $(8.5,8)$, very close to Naples in 2019 (see Figure \ref{fig:RVFAttractors} below for a confirmation).
In the HL quadrant, for a relatively low density, arrows point to upper right, i.e. to a dynamics of concentrated urban growth; on the contrary, for high density, the direction is mainly toward the left, which suggests the presence of urban sprawl.
In a large part of the LH quadrant there are no significant arrows. The only significant ones generally point downward and on the right, indicating a possible pattern of agglomeration, i.e. populations tend to cluster in specific municipalities while their LLA become less dense. Another attractor around $(2.5,5)$ is present (see Figure \ref{fig:RVFAttractors}).
Finally, in the middle of the LL quadrant, arrows clearly point downwards to the left, i.e. to a depopulation in the municipalities and their LLAs. Differently, in the part of quadrant with high-density LLA the dynamics points to LH quadrant, while below the bisector the upward trend seems to be dominant. Nonetheless, the complex pattern, an attractor around $(0,3)$ is present (see Figure \ref{fig:RVFAttractors}). The non-parametric Moran does not show any appreciable change in the quadrant over time.help the design of policies contrasting the socio-economic decline of some Italian areas.

\subsection{The forecast of demographic trends in Italian municipalities}\label{subsec:attractors}

\begin{figure}[!htbp]
	\centering
	\begin{subfigure}[b]{0.58\textwidth}
		\centering
		\includegraphics[width=\textwidth]{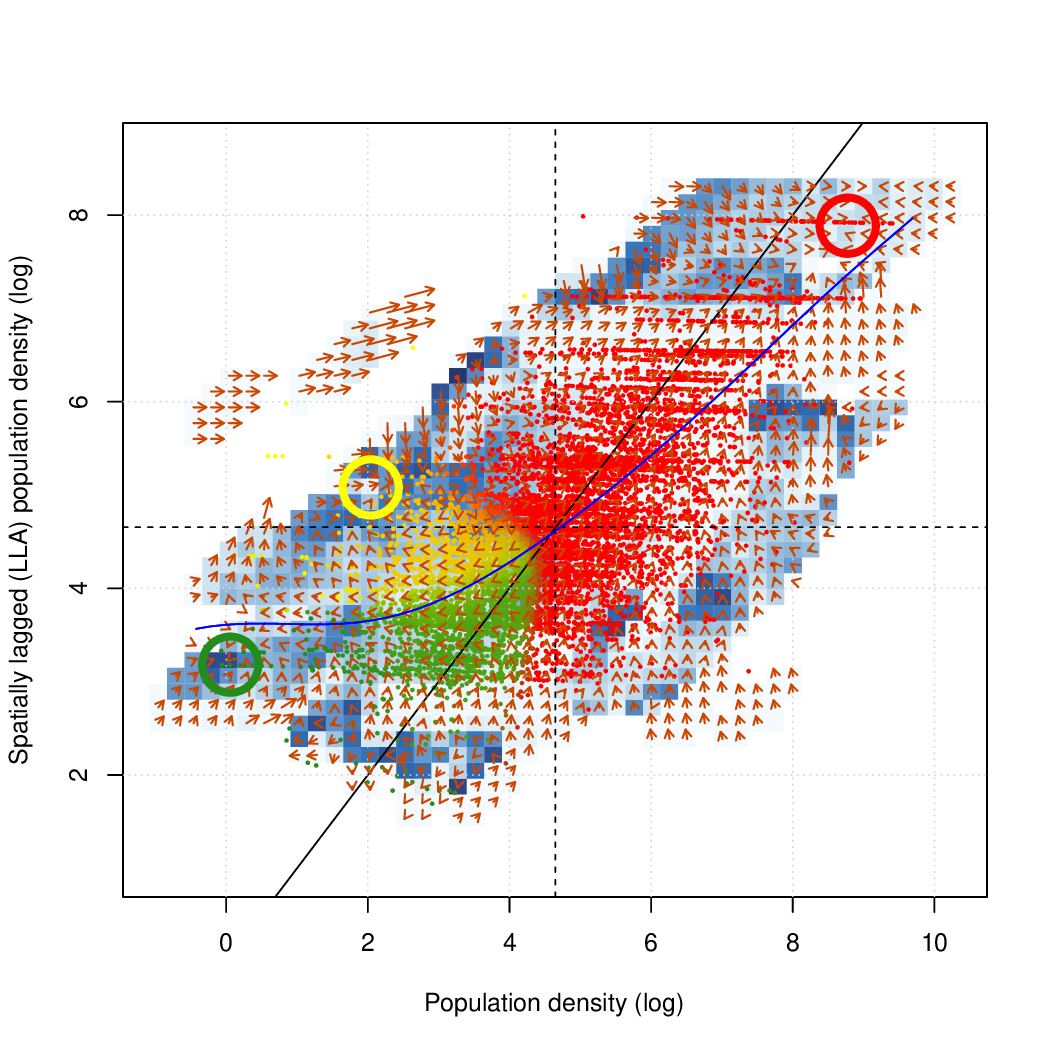}
   \caption{The three attractors and the observations in 2019 coloured according to their probability to belong to one of the three attractors.}
  \label{fig:RVFAttractors}	
	\end{subfigure}
\vspace{0.05cm}
 \begin{subfigure}[b]{0.41\textwidth}
     \includegraphics[width=\textwidth]{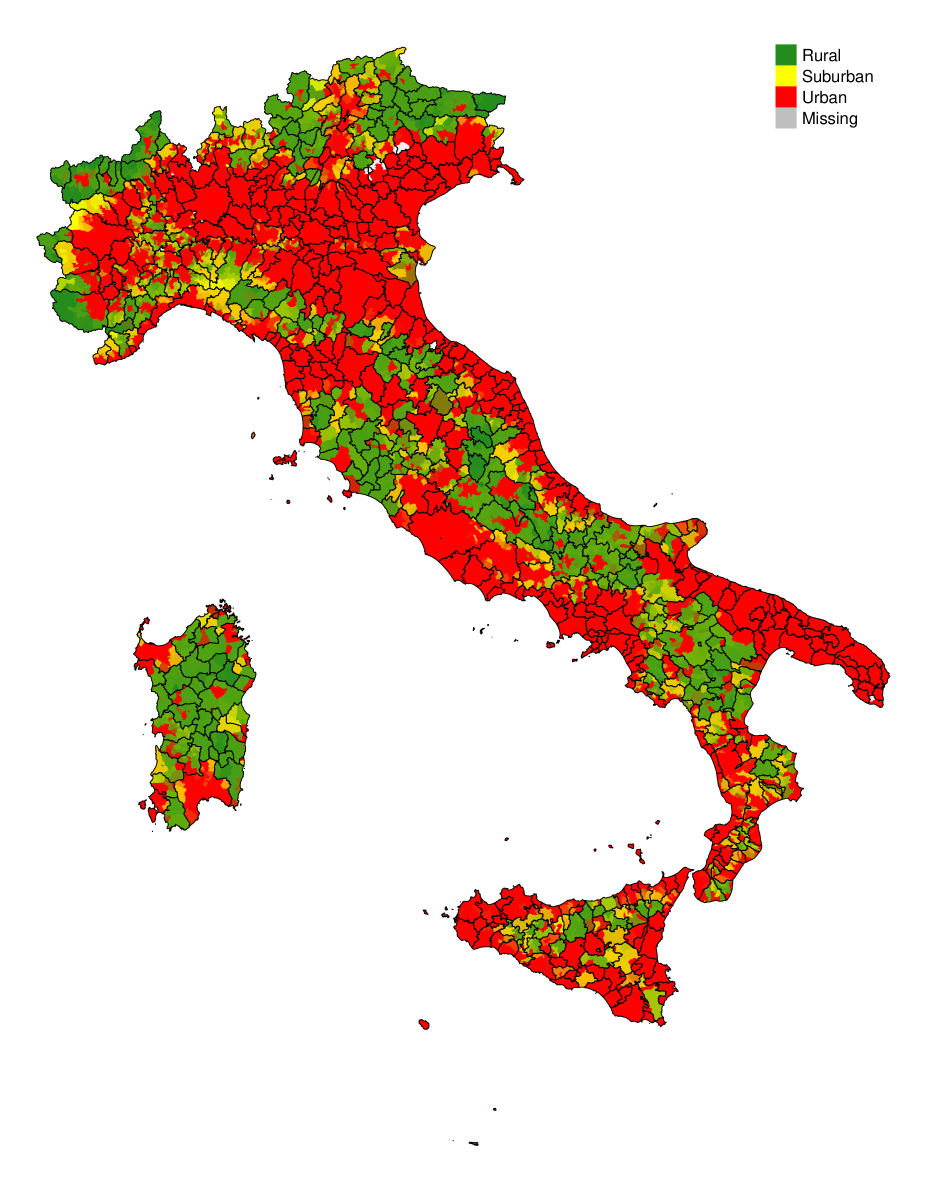}
     \caption{The map of Italian municipalities coloured on the base of their probability to belong to one of the three attractors.}
     \label{fig:mapAttractors}
 \end{subfigure}
\caption{Panel on the left: the estimate of RVF of Eq. \eqref{eq:RVF} in the Moran space for the period 1984-2019. Red arrows are the directions in each evaluation point (the length of each arrow has been divided by a factor 5 for graphical purposes). Estimated variance of the arrows' direction is reported in blue (the darker the colour the higher the variance). The yellow, red and green circles point to the locus of the three attractors. Observations of 2019 are coloured with shades of yellow, red and green according to their probability to belong to one of the three attractors. Blue line is the estimated non-parametric Moran in 2019. Panel on the right: the map of Italian municipalities coloured with shades of yellow, red and green according to their probability to belong to one of the three attractors. \\ \textit{Source: our elaborations on ISTAT data (Italian National Institute of Statistics).}
}
\end{figure}

From the estimate of RVF it is possible to forecast the dynamics of Italian municipal population density, which heuristically means to simulate the trajectory of each municipality taking 2019 as the starting point on the Moran space on the base of the estimated arrows (see Appendix \ref{app:contForw}).\footnote{A similar methodology is used in \cite{fiaschi2018spatial}.} The forecast should be interpreted as the future evolution of actual demographic trend in the Italian municipalities under the assumption that RVF is constant along the forecast horizon. In the forecast, we use all the estimated directions independently of their statistical significance, and we make the inference on forecast by a bootstrap procedure (see Appendix \ref{app:inference RVF}). 
A first goal of the forecast is to discover possible attractors in the Moran space, i.e. the areas where municipalities and their LLAs tend to converge in the long run and their basin of attraction. We find three attractors, which are reported in yellow, red and green circles in Figures \ref{fig:RVFAttractors} and \ref{fig:mapAttractors}.\footnote{Missing values in the map are due to change in the municipal definition from 1984 to 2019.} The red circle around (8.5,8) in the HH quadrant points to an \textit{urban attractor} (UA), the yellow circle around (2.5,5) in the LH quadrant to a \textit{suburban attractor} (SA), and the green circle around (0,3) in the LL quadrant to a \textit{rural attractor} (RA). The three attractors corresponds to three prototypical settlement systems.

The basin of attractions are identified in Figure \ref{fig:RVFAttractors} by colouring the observations of 2019 according to their probability to belong to one of the three attractors.
The basin of the UA is composed of both the HH and HL quadrants as well as some parts of LH  and LL quadrants; the basin of the SA expands both in the LH and LL quadrants above the non-parametric Moran; finally, the basin of the RA is totally in LL and LH quadrant below the non-parametric Moran. In particular, the UA accounts for 65\% of municipalities and 93\% of population in 2019, the SA accounts for 14\% of municipalities and 2\% of the total population, and the RA refers to 21\% of municipalities and 5\% of population (see Table \ref{tab:attractors}). Therefore, the 21\% of municipalities is at risk of depopulation, corresponding to 5\% of Italian population. Another 14\% of municipalities is subject to urban sprawling, corresponding to 2\% of Italian population.

\begin{table}[!htbp]
\centering
\begin{tabular}{lccc} \hline \hline \\[-0.3cm]
                & \multicolumn{1}{l}{Urban attractor} & \multicolumn{1}{l}{Suburban attractor} & \multicolumn{1}{l}{Rural attractor} \\ \hline \\[-0.35cm]
Municipality & $65\%$                     & $14\% $                   & $21\%$                  \\
 & \footnotesize{$(62.5\%-74.5\%)$}  & \footnotesize{$(1.1\%-33.5\%)$} &  \footnotesize{$(1.5\%-30\%)$}\\ \\[-0.35cm]
 Population     & $93\%$                   & $2\% $                    & $5\% $ \\
 & \footnotesize{$(92.1\%-94.7\%)$}  & \footnotesize{$(0.1\%-6.9\%)$} &  \footnotesize{$(0.1\%-6.7\%)$}\\ 
	                 \hline \hline
\end{tabular}
\caption{Percentage of municipalities and population in 2019 in the urban attractor (UA), suburban attractor (SA) and rural attractor (RA) forecasted by the estimated RVF reported in Figure \ref{fig:RVFAttractors}. Confidence bands at 90\% are reported in parenthesis. \\ \textit{Source: our elaborations on ISTAT data (Italian National Institute of Statistics).}
}
\label{tab:attractors}
\end{table}

\section{RVF as a tool for the National Strategy for Inner Areas (SNAI) \label{sec:innerAreas}}

In this section we will discuss how the estimated RVF can be used to help the design of policies contrasting the socio-economic decline of some Italian areas. In particular, since 2012 Italy is developing a \textit{National Strategy for Inner Areas} (\textit{Strategia Nazionale per le Aree Interne}, SNAI), where the inner areas are defined in \cite{barca2014}. The main idea is to exploit the socio-economic potential of these inner areas, characterized by a considerable distance from places providing essential services  (\textit{hubs}) but, at the same time, displaying remarkable environmental and cultural resources and with a history of human habitation spanning centuries.
In fact, a significant proportion of inner areas are experiencing depopulation, a decline in local public and private services, and degradation of the cultural and landscape heritage. ``Summarising the ultimate objective and guiding light of the strategy is to reverse and improve demographic trends [...]'' \citep[p. 8]{barca2014} .
 
\begin{figure}[!htbp]
	\centering
	\begin{subfigure}[t]{0.70\textwidth}
	 \includegraphics[width=\textwidth]{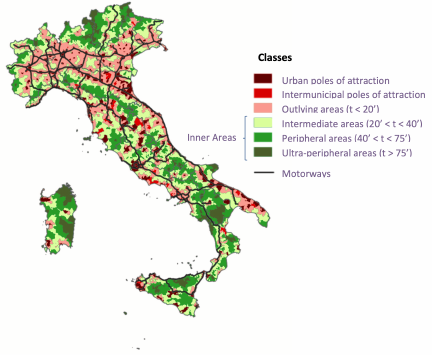}
\caption{Figures III.2 in taken from \cite[p. 26]{barca2014}.}
\label{fig:Barca2014}	
	\end{subfigure}
	\begin{subfigure}[t]{0.47\textwidth}
		\centering
	\includegraphics[width=\textwidth]{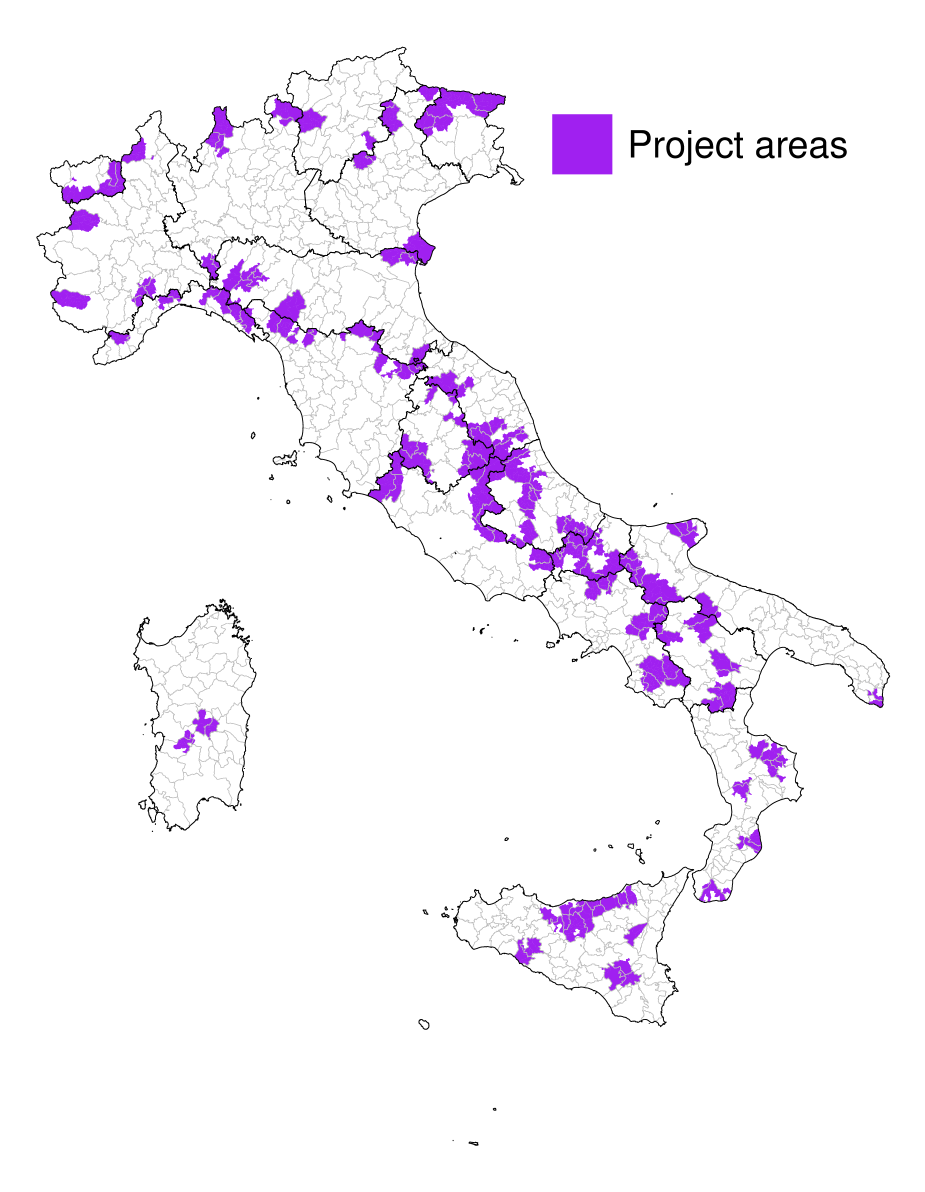}
		\caption{The map of project areas financed in 2014-2020. \\ \textit{Source: 
			Dipartimento per le politiche di coesione, Presidenza del Consiglio dei Ministri.}}
    \label{fig:mapFinanced}
	\end{subfigure}
	\begin{subfigure}[t]{0.47\textwidth}
		\centering
		\includegraphics[width=\textwidth]{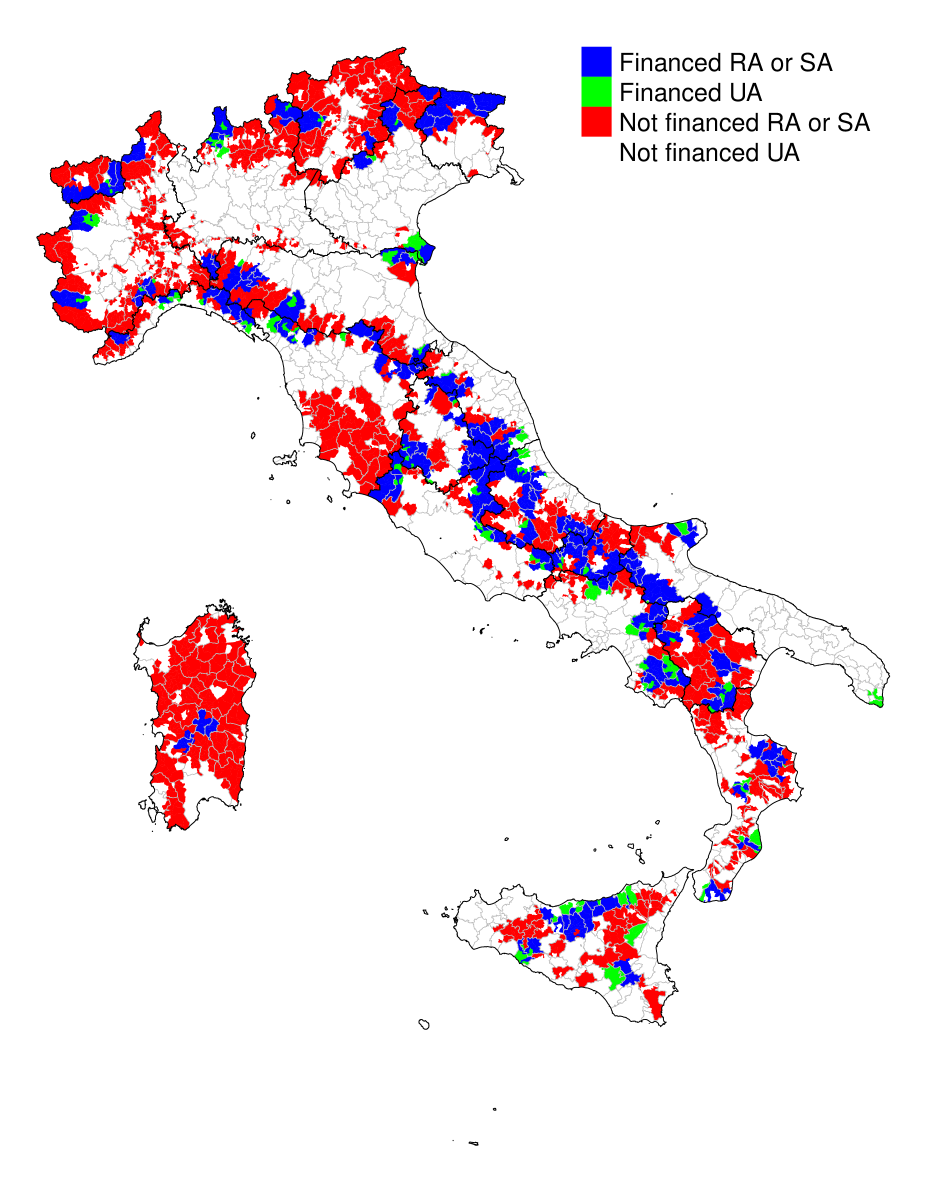}
		\caption{The map of urban areas (UA), suburban areas (SA) and rural areas (RA) financed and not financed in 2014-2020.\\ \textit{Source: our elaborations on ISTAT data (Italian National Institute of Statistics) and Dipartimento per le politiche di coesione, Presidenza del Consiglio dei Ministri.}}
  \label{fig:mapRASA}
	\end{subfigure}
\label{fig:InnerAreas}
\end{figure}

In particular, according to \cite{barca2014}, Italian municipalities can be divided into: ``\textit{Urban poles of attraction}'' and ``\textit{Intermunicipal poles of attraction}'' which are either single municipalities or a group of neighbouring municipalities able to provide a full range of secondary education, at least one grade 1 emergency care hospital, and at least one Silver category railway station; ``\textit{Outlying areas}'', ``\textit{Intermediate areas}'', ``\textit{Peripheral areas}'' and ``\textit{Ultra-peripheral areas}'' which do not have all these essential services and are located within a distance of 20, 40, 75 or more minutes from the nearest hub, respectively (see Figure \ref{fig:Barca2014}). Municipalities belonging to Intermediate, Peripheral and Ultra-peripheral areas are labelled as \textit{Inner Areas} by \cite{barca2014}.

Although in the definition of Inner Areas no demographic criteria have been applied, since the late 1970s both ``Peripheral areas'' and ``Ultra-peripheral areas'' exhibited negative demographic trends. As a result, these areas are the focus of targeted funding initiatives devoted to increase the provision of essential public services. 
Given the principle of concentration of the EU cohesion policy, SNAI  does not operate on all the Inner Areas' municipalities, but it concentrates in appropriately selected project areas, which group neighbouring municipalities. In the selection, more than 100 socio-economic factors were considered, among which demographic composition and trends, agricultural land use, education, transportation and healthcare services. Over the period 2014-2020, 72 project areas were selected  covering 1,060 municipalities and about 2 millions of inhabitants.\footnote{See \url{https://politichecoesione.governo.it/it/strategie-tematiche-e-territoriali/strategie-territoriali/strategia-nazionale-aree-interne-snai/le-aree-interne-2014-2020/} for a complete list.}

In the selection of the project areas, one of the criteria is based on the variation in total population between 1971 and 2011, while another on the variation of foreign population between 2001 and 2011. In this respect, the estimated RVF can help the identification of such areas by providing a more reliable forecasting of the long-term demographic pattern for each municipality, which adds to the demographic history of each municipality also the past dynamics of its neighbours. In particular, the membership of one municipality to a specific basin of attraction provides key information on its expected population trajectory, which can be taken as an identification of the candidate target areas in their demographic dimension \citep[p. 16]{barca2014}. These ``intervention-free'' trajectory is calculated considering the municipality not as an isolated entity, but taking into account the evolution of the entire its LLA, which synthesizes the information of where people live and where job opportunities and services are located.

A comparison between Figures \ref{fig:Barca2014} and \ref{fig:mapFinanced} highlights how only a (small) fraction of total inner areas is also a project area as outcome of the principle of concentration of the EU cohesion policy and a binding budget constraint.
Figure \ref{fig:mapRASA} shows that almost all the project areas are composed by suburban and rural municipalities (SA and RA, respectively) as identified by the RVF in terms of their basin of attraction. Only few urban municipalities (UA) result members of project areas, and in the most of the cases this can be explained with the principle of geographic contiguity used in the identification of project areas. The only exception is the southern part of Apulia (sud Salento). Such a principle is however not fully respected when there exist regional boundaries, as it is evident considering the southern municipalities of Tuscany and Valle d'Aosta, the northern municipalities of Liguria and Veneto, as well as the majority of municipalities in Basilicata.

\section{Conclusions}\label{sec:conclusion}

In this paper, we analyse the time evolution of the distribution of population density across Italian municipalities, employing a methodology that is able to account explicitly for spatial dependence. The estimation of the RVF in the Moran space is based on a weighted average of all the spatial transitions observed over the period, where the weights are derived by non-parametric kernel density estimation.
The main advantage of the proposed methodology is its ability to mitigate the curse of dimensionality in non-parametric estimation, while keeping the dependence on space explicit. The estimate of RVF allows the identification of the long-run behaviour as well as 
the spatial attractors. 

When we apply the RVF to estimate the spatio-temporal dynamics of the population density of Italian municipalities, we find evidence of the presence of three attractors. The identification of such attractors provides a more reliable forecasting of the long-term demographic pattern for each municipality, which adds to the demographic history of each municipality also the past dynamics of its neighbours.
On the base of forecasted demographic trajectories we were able to partition municipalities into three main categories: the 21\% of municipalities is at risk of depopulation, corresponding to 5\% of Italian population, while another 14\% of municipalities is subject to urban sprawling, corresponding to 2\% of Italian population. The remaining municipalities and population should display a process of agglomeration and growth urban concentration.

We discuss how our methodology validates the identification of project areas made in \textit{National Strategy for Inner Areas}. We find evidence of a not sufficient funding for municipalities expected to be rural (or suburban) in the future and of a border effect in the actual funding strategy, likely due to the design of the decision process, which demands to NUTS-2 regions the proposals of the areas to be funded.\footnote{See \url{https://politichecoesione.governo.it/it/strategie-tematiche-e-territoriali/strategie-territoriali/strategia-nazionale-aree-interne-snai/le-aree-interne-2014-2020/} for the details on the design of the intervention.} Such border effect could be a severe source of inefficiency in presence of positive spatial spillovers.
Finally, we advocate that, when the main interest of analysis is the estimate of future tendencies but the set of information is very limited, the proposed methodology can be applied to the study other socio-economic phenomena characterized by both spatial and temporal dynamics, as personal income, poverty, and  social exclusion.

\bigskip

\noindent\textbf{Acknowledgements}: The authors have been supported by the Italian Ministry of University and Research (MIUR), in the framework of PRIN project 2017FKHBA8 001 (The Time-Space Evolution of Economic Activities: Mathematical Models and Empirical Applications).
Davide Fiaschi has been supported by the University of Pisa, in the framework of the PRA Project PRA\_2022\_86 (Mobilità di persone e merci nell'Europa).\\
\noindent\textbf{Declaration of interest}: The authors declare that they have no known competing financial interests or personal relationships that could have appeared to influence the work reported in this paper.\\
\noindent\textbf{Code availability}: All the codes are free to use and available at the link: \url{https://github.com/PRINtimeSpaceEconAct/RVF_estimate}.

\clearpage

\bibliographystyle{apalike}

\bibliography{biblio}

\newpage

\appendix

\section{Technical details on the estimation of RVF and forecast}\label{app:RVF_flavour}

In this appendix we detail the technicalities behind the estimation of a RVF (a simplified exposition was already given in Section \ref{sec:timeSpace}) and its use to compute the forecast of spatial distribution.

\subsection{The adaptive kernel estimator with full covariance matrix}

The RVF reported in Figure \ref{fig:RVF_84_19} is based on weights computed from the joint density of $(Y^{t_0},WY^{t_0})$, estimated using an \textit{adaptive kernel method} and keeping into account the covariance of observations, as suggested by \cite{silverman1986density}.
In particular: 
\begin{enumerate}
	\item run the \emph{pilot density estimate}, with bandwidth $h$ in evaluation point $\mathbf{z} \in \RR^2$, defined as:
\begin{equation*}
   \tilde{f}_{Y^{t_0},WY^{t_0}}(\mathbf{z}):=\frac{\text{(det}S)^{-1/2}}{Nh^2}\sum_{j=1}^N k\left(\dfrac{(\mathbf{z}-\mathbf{z}^{t_0}_{j})^T S^{-1}(\mathbf{z}-\mathbf{z}^{t_0}_{j})}{h^2}\right),
\end{equation*}
where $k(\mathbf{z}^T\mathbf{z}) = K(\mathbf{z})$, $K$ is a kernel function over $\RR^2$ and $S$ is the sample covariance matrix of observations. 
 \item Calculate the \emph{local bandwidth factors} $\lambda_j$ as 
\begin{equation*}
\lambda_j := \left(\frac{\tilde{f}_{Y^{t_0},WY^{t_0}}(\mathbf{z}^{t_0}_j)}{g}\right)^{-\alpha},
\end{equation*}
where $g$ is the geometric mean of $\tilde{f}_{Y^{t_0},WY^{t_0}}(\mathbf{z}^{t_0}_j)$, i.e.
\begin{equation*}
g := \exp \left(  \frac{1}{N}\sum_{j=1}^N \tilde{f}_{Y^{t_0},WY^{t_0}}(\mathbf{z}^{t_0}_j)  \right).
\end{equation*}
The parameter $\alpha \in (0,1)$ is the sensitivity parameter to local density, i.e. the strength of adaptation of the estimate to local density of observations ($\alpha=0$ no adaptation) \citep{silverman1986density}. 
\item Finally, run the \emph{adaptive kernel density estimator}: 
\begin{equation*}
\hat{f}^a_{Y^{t_0},WY^{t_0}}(\mathbf{z}):=\frac{\text{(det}S)^{-1/2}}{N}\sum_{j=1}^N \frac{1}{h^2\lambda_j^2}k\left(\dfrac{(\mathbf{z}-\mathbf{z}^{t_0}_{j})^T S^{-1}(\mathbf{z}-\mathbf{z}^{t_0}_{j})}{h^2\lambda_j^2}\right).
\end{equation*}
\end{enumerate}

In our case the adaptive approach is really effective since observations are spread very far apart from the median, with regions displaying a high number of observations and others with only few observations. 
Therefore, calculate the \textit{RVF adaptive estimator} $\widehat{\mathbf{RVF}}^a_\tau$ as
\begin{equation}
\widehat{\mathbf{RVF}}^a_\tau(\mathbf{z}):=\sum_{j=1}^N \omega_j^{a,t_0}(\mathbf{z})\boldsymbol{\delta}_j^\tau,
    \label{eq:RVFApp}
\end{equation}
where:
\begin{equation*}
    \omega_j^{a,t_0}(\mathbf{z}):= \dfrac{\dfrac{\text{(det}S)^{-1/2}}{Nh^2\lambda_j^2} k\left(\dfrac{(\mathbf{z}-\mathbf{z}^{t_0}_{j})^T S^{-1}(\mathbf{z}-\mathbf{z}^{t_0}_{j})}{h^2\lambda_j^2}\right)}{\hat{f}^a_{Y^{t_0},WY^{t_0}}(\mathbf{z})}.
\end{equation*}

\subsection{The choice of optimal estimation parameters $h$ and $\alpha$}

The choice of parameters $\alpha$ and $h$ of Eq. \eqref{eq:RVFApp} is made by minimizing the mean square error (MSE) of the estimate for a set of candidate couples.
In particular, for each possible couple of $(\alpha, h)$ estimate the adaptive RVF, $\widehat{\mathbf{RVF}}^a_\tau(\mathbf{z})$; then, use the estimated RVF, given the initial observations, to predict their values in the final year, i.e.
\begin{equation}\label{eq:zt0tauhat}
\hat{\mathbf{z}}_j^{t_{0}+\tau} := \Phi^{\tau}_{\widehat{\mathbf{RVF}}^a_\tau}({\mathbf{z}}_j^{t_{0}}),
\end{equation}
where we refer to Section \ref{app:contForw} for the details of function $\Phi^{\tau}_{\widehat{\mathbf{RVF}}^a_\tau}$ which predicts the observations in the last year given the ones in the first year.
Finally, the goodness of the estimation is measured by the mean of the square distance between the predicted observations $\hat{\mathbf{z}}_j^{t_{0}+\tau}$ and the actual observation at the final year ${\mathbf{z}}_j^{t_{0}+\tau}$, i.e.:
\begin{equation*}
MSE(\alpha,h) := \frac{1}{N}\sum_{j=1}^N \norm{\hat{\mathbf{z}}_j^{t_{0}+\tau} - {\mathbf{z}}_j^{t_{0}+\tau}}^2.
\end{equation*} 
In practice, the optimal couple of $(\alpha,h)$, i.e. the one minimizing $MSE(\alpha,h)$, is searched over a finite grid, which in our case consists of all their possible combinations in the sets
$h = \{0.04, 0.06, 0.08, 0.11, 0.15, 0.21, 0.29,0.41, 0.57, 0.80\}$ and $\alpha = \{0, 0.0005, 0.0012, 0.0028, 0.0067, 0.0158, 0.0375, 0.0889, 0.2108, 0.5\}$ for a total of 100 estimated RVFs. The choice of the grid follows the methodology suggested by \cite{bowman1997applied}. The optimal combination results $(\alpha^\star,h^\star)=(0.0067,0.21)$.
 
\subsection{The forecast via RVF in continuos time}\label{app:contForw}

Differently from \cite{fiaschi2018spatial}, forecast is made in continuos time by function $\Phi^{\tau}_{\widehat{\mathbf{RVF}}^a_\tau}$ in Eq. \eqref{eq:zt0tauhat}.
Given the vector field $F:\RR^2 \to \RR^2$ and an observation $j$ at the initial year $\mathbf{z}^{t_0}_j \in \RR^2$, first calculate the trajectory that observation $j$, $\mathbf{z}^{t_0}_j$, will have following the \textit{vector field} $F$. In particular, introduce the differential equation for the variable $p^{z^{t_0}_j}$ as follows:
\begin{equation}\label{eq:ODEContinuousForward}
\begin{cases}
\dot{p}^{z^{t_0}_j} = F(p^{z^{t_0}_j}); \text{and}\\
p^{z^{t_0}_j}(t_0) = z^{t_0}_j.
\end{cases}
\end{equation}
Roughly speaking the curve $p^{z^{t_0}_j}:[t_0,+\infty)\to \RR^2$ represents the trajectory of observation $j$, which starts at $z^{t_0}_j$ at time $t_0$ and follows the direction prescribed by $F$; $\Phi^{\tau}_{F}(z^{t_0}_j)$ is therefore defined as:
\begin{equation}
\Phi^{\tau}_{F}(z^{t_0}_j) := p^{z^{t_0}_j}(t_0+\tau)\in \RR^2.
\end{equation}
In other words, $\Phi^{\tau}_{F}(z^{t_0}_j)$ is the value at time $t_0+\tau$ of an observation $j$ whose value at time $t_0$ was $z^{t_0}_j$, and has followed the arrows prescribed by the vector field $F$. In our model, $\hat{\mathbf{z}}_j^{t_{0}+\tau}$ of Eq. \eqref{eq:zt0tauhat} represents the position in Moran space that observation $j$ will reach following the estimated RVF for $\tau$ periods.

In practice, to numerically solve Eq. \eqref{eq:ODEContinuousForward} we have to compute $F$ in every point of the trajectory of $p^{z^{t_0}_j}$. However, in our case the vector field to predict the dynamic $\widehat{\mathbf{RVF}}^a_\tau$ can only be estimated on a finite set of evaluation points.
Therefore, we rely on an interpolation procedure: given a finite regular grid of evaluation points where $\widehat{\mathbf{RVF}}^a_\tau$ has been estimated, and for each point $\mathbf{z}$ for which we have to evaluate $\widehat{\mathbf{RVF}}^a_\tau(\mathbf{z})$ (assuming $\mathbf{z}$ itself is not one of the evaluation points), first consider the four evaluation points defining the corners of the square where $\mathbf{z}$ belongs to. Then, the value of $\widehat{\mathbf{RVF}}^a_\tau(\mathbf{z})$ is approximated by linearly interpolating the values of the estimated RVF in the four corners. This produces an approximated estimated RVF in every point $\mathbf{z} \in \RR^2$, thus allowing to numerically solve Eq. \eqref{eq:ODEContinuousForward}.

\subsection{The inference in RVF \label{app:inference RVF}}

The inference in RVF is made by a bootstrap procedure which considers the transitions as the unit to be sampled with repetition. This is equivalent to a block bootstrap procedure, which takes into account the spatial correlation of observations. In particular, assume to observe $N$ transitions in the Moran scatter plot; draw a bootstrap sample of size $N$; estimate the RVF for the bootstrap sample; repeat this procedure $B$ times. The outcome is a collection of $B$ RVFs, by which we can infer the significance of estimated direction in each evaluation point. In the forecast, the collection is also used to infer with which probability the trajectory starting from a given evaluation point is converging to a specific attractor.

\section{Change in LLA definition}\label{app:changeLLA}

\begin{figure}[!htbp]
	\centering
	\includegraphics[width=1\textwidth]{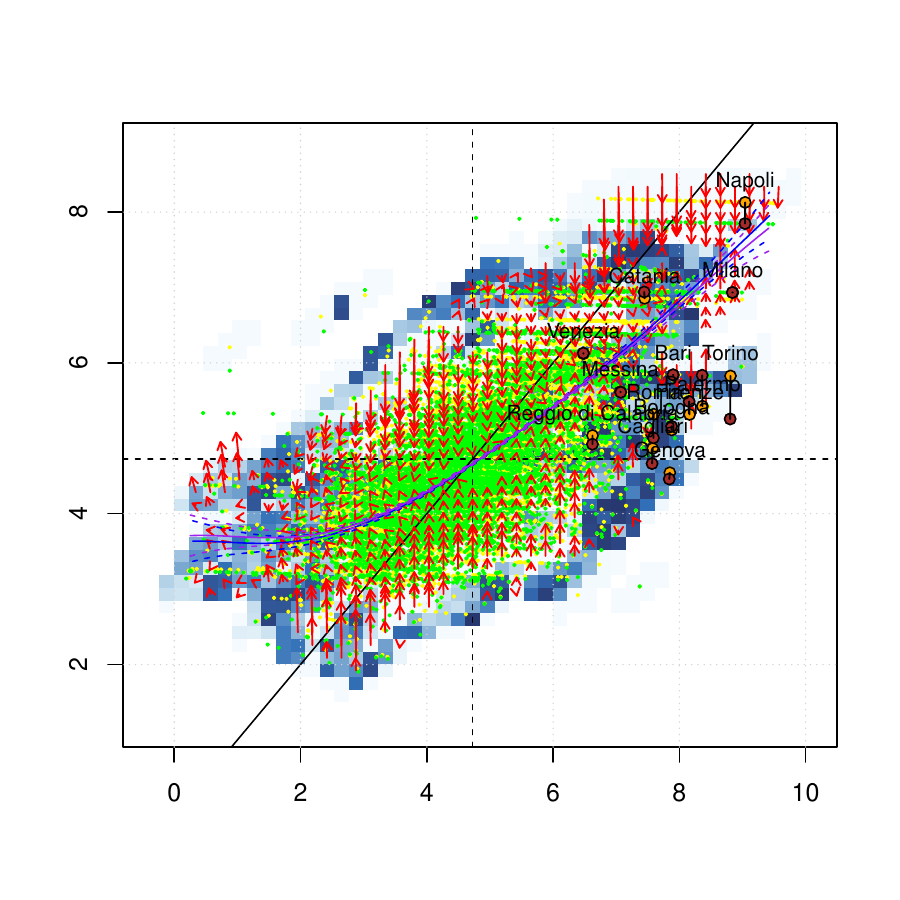}
	\caption{The estimate of RVF of Eq. \eqref{eq:RVF} in the Moran space for the period 2001-2002. Red arrows are the directions statistically significant at 5\% level (the length of each arrow has been divided by a factor 5 for graphical purposes).
		Estimated variance of the arrows' direction is reported in blue (the darker the colour the higher the variance). Observations in 2001 and 2002 are coloured in yellow and green, while metropolitan cities in orange and brown, respectively. Blue and purple lines are the non-parametric estimate of Morans and their 95\% confidence bands in 2001 and 2002.
		\\ \textit{Source: our elaborations on ISTAT data (Italian National Institute of Statistics).}}
	\label{fig:RVF_01_02}
\end{figure}

The RVF estimation has been carried out considering two different definitions of LLA for 1984 and 2019. As explained in Section \ref{sec:data} we use the LLA definition of 2001 for the years 1984-2001 and that of 2011 for the years 2002-2019. To understand how much our estimations depend on this change in Figure \ref{fig:RVF_01_02} we report the RVF estimated only on the adjacent years 2001 and 2002. The short time span should minimize the change in municipal population density and, therefore, to help in  isolating the effects of the change in LLA definition on estimated RVF.
In Figure \ref{fig:RVF_01_02}, as expected, almost all the arrows are vertical (either upwards or downwards) confirming that the change in municipal population density between 2001 and 2002 is negligible. Moreover, all the arrows point to the non-parametric Moran, i.e. the change in the LLA definition has favoured the convergence toward non-parametric Moran described in Section \ref{sec:timeSpace}.

\section{List of metropolitan cities \label{app:listMetroCities}}

Table \ref{tab:listMetroCities} report the population density and the total population in 1984 and 2019 for the 14 Italian metropolitan cities.
% latex table generated in R 4.1.0 by xtable 1.8-4 package
% Mon Feb  6 09:56:33 2023
\begin{table}[!htbp]
\centering
\begin{tabular}{lrrrr}
  \hline
  \hline
 Municipality & Density  & Population  & Density  & Population \\ 
  & in 1984 & in 1984 &  in 2019 & in 2019\\ 
  \hline
 Rome & 7.70 & 2844903 & 7.69 & 2820219 \\ 
 Milan & 9.06 & 1560155 & 8.95 & 1395980 \\ 
 Naples & 9.23 & 1207337 & 8.99 & 954318 \\ 
 Turin & 9.04 & 1097355 & 8.80 & 860793 \\ 
 Palermo & 8.40 & 711194 & 8.31 & 652720 \\ 
 Genoa & 8.05 & 748634 & 7.77 & 569184 \\ 
 Bologna & 8.06 & 447893 & 7.93 & 393248 \\ 
 Florence & 8.37 & 442345 & 8.19 & 369885 \\ 
 Bari & 8.05 & 368772 & 7.90 & 316491 \\ 
 Catania & 7.64 & 380564 & 7.40 & 297752 \\ 
 Venice & 6.70 & 339883 & 6.44 & 259961 \\ 
 Messina & 7.08 & 254951 & 6.98 & 229280 \\ 
 Reggio di Calabria & 6.61 & 177807 & 6.60 & 176299 \\ 
 Cagliari & 7.93 & 236165 & 7.49 & 151504 \\ 
   \hline
   \hline
\end{tabular}
\caption{The population density and the total population of the 14 Italian metropolitan cities in 1984 and 2019. \\ \textit{Source: ISTAT (Italian National Institute of Statistics).}}
\label{tab:listMetroCities}
\end{table}

\end{document}